\documentclass[%
 aip,
 amsmath,amssymb,
 reprint,%
]{revtex4-1}
\usepackage[none]{hyphenat}

\usepackage{graphicx}
\usepackage{dcolumn}
\usepackage{bm}
\usepackage[utf8]{inputenc}
\usepackage[T1]{fontenc}
\usepackage{mathptmx}
\usepackage{etoolbox}
\usepackage{hyperref}
\usepackage{placeins}

\usepackage{booktabs}
\usepackage{multirow}

\makeatletter
\def\@email#1#2{%
 \endgroup
 \patchcmd{\titleblock@produce}
  {\frontmatter@RRAPformat}
  {\frontmatter@RRAPformat{\produce@RRAP{*#1\href{mailto:#2}{#2}}}\frontmatter@RRAPformat}
  {}{}
}%
\makeatother

\begin{document}

\title{Interpretable Geometry Sensitivity for Inverse Design of Integrated Photonics}

\author{Junho Park}
\thanks{These authors contributed equally to this work.}
\affiliation{Department of Electrical and Computer Engineering, University of Michigan, Ann Arbor}
\email{junhop@umich.edu}

\author{Taehan Kim}
\thanks{These authors contributed equally to this work.}
\affiliation{College of Computing, Data Science, and Society, University of California, Berkeley}

\author{Mohammad Ali}
\affiliation{Department of Electrical and Computer Engineering, University of Michigan, Ann Arbor}

\author{Di Liang}
\affiliation{Department of Electrical and Computer Engineering, University of Michigan, Ann Arbor}

\date{\today}

\begin{abstract}
As an increasingly powerful technique in integrated photonics, inverse design uses optimization algorithms to automatically create compact, high-performance photonic structures, often yielding non-intuitive layouts far more compact than conventional designs.\cite{jensen2011topology, molesky2018inverse} While adjoint-based inverse design is a prominent optimization method, the resulting free-form layouts are difficult to interpret or diagnose under fabrication variability, even for experienced photonic device designers.\cite{lalau2013adjoint, su2019nanophotonicinversedesignspins, piggott2015inverse} We present an experimentally validated interpretability workflow that produces pixel-level sensitivity maps directly on the binary mask of an inverse-designed device. Using wavelength-division demultiplexers (WDMs) at 1310/1550~nm as examples, we train a lightweight convolutional surrogate to regress figures of merit (FoMs) and apply Integrated Gradients (IG) to attribute predicted transmission to individual pixels. We demonstrate that high-attribution hotspots correspond to physically meaningful substructures, such as splitter hubs and high-curvature edges. Experimental results show that controlled perturbations in these high-sensitivity regions result in up to an $\sim$11x higher excess insertion loss compared to perturbations in non-sensitive regions, consistent with full-wave simulations. This approach adds a practical explainability layer to existing pipelines, offering a clear pathway for foundry-compatible design-rule checking and fabrication-aware constraint allocation without modifying the underlying electromagnetic solver.
\end{abstract}

\maketitle

\section{Introduction}
Adjoint-based inverse design has enabled compact nanophotonic devices for filtering, coupling, and routing by optimizing over high-dimensional design spaces (e.g., pixelated topology variables).\cite{lalau2013adjoint,molesky2018inverse,su2019nanophotonicinversedesignspins} In integrated photonics practice, however, the geometric complexity that enables performance can also hinder adoption: inverse-designed masks often contain small, non-intuitive features that are difficult to interpret, enforce with fabrication constraints, and debug when measured performance deviates from simulation. \cite{hammond2021photonic, piggott2017fabrication} This lack of interpretability directly impacts design-rule checking, mask quality control, targeted Scanning Electron Microscopy (SEM) inspection, and principled simplification/cleanup for foundry integration.\cite{hammond2021photonic}

A core obstacle is that adjoint optimization provides gradients with respect to design variables,\cite{miller2012photonic} but does not directly yield a human-interpretable map of which {\em finalized} geometric features are most responsible for key metrics at specific operating wavelengths. Moreover, when a fabricated device underperforms, designers frequently resort to global re-optimization or ad-hoc edits, rather than localized, sensitivity-informed modifications.

In parallel, explainable artificial intelligence (XAI) methods such as Local Interpretable Model-agnostic Explanations (LIME) \cite{LIME} and Integrated Gradients (IG) \cite{IG} have been developed to attribute model predictions to input features. These build upon earlier efforts in computer vision to produce saliency maps that visualize the internal representations of deep networks.\cite{simonyan2013deep} Among these methods, IG is particularly attractive for photonic design layouts because it is gradient-based, preserves geometric structure, and can provide pixel-level attributions on the original input mask. These properties align naturally with binary photonic masks, motivating an interpretability layer that maps predicted optical metrics back to localized geometry.

Complementing these XAI frameworks, neural network based surrogate models serve as efficient estimators for the complex, high-dimensional mapping between photonic geometry and electromagnetic response.\cite{peurifoy2018nanophotonic, wiecha2021deep, liu2018training} By providing a differentiable approximation of full-wave Maxwell solvers, these surrogates enable rapid exploration of design landscapes that are otherwise computationally intractable.

Our work suggests that a differentiable surrogate trained on diverse inverse-designed masks can be used as an ``analysis instrument'': IG applied to the surrogate model yields a practical, mask-aligned sensitivity map that (i) highlights physically meaningful substructures (e.g., splitter hubs, tapers, sharp corners) and (ii) predicts which localized edits are likely to most affect transmission. We demonstrate the method on inverse-designed two-channel WDMs targeting O-band (1310~nm) and C-band (1550~nm), and we validate the pixel-level sensitivity map using matched-area region perturbations with FDTD simulation and experimental validation.

\begin{figure*}[t]
\centering
\includegraphics[width=\textwidth]{"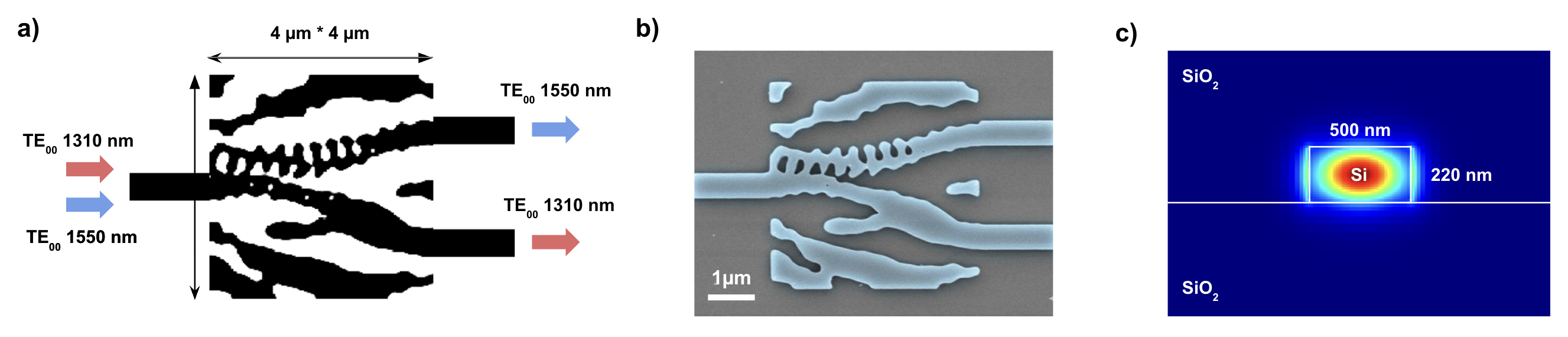"}
\caption{\label{fig:device}
(a) Inverse-designed two-channel WDM (1310/1550~nm) (b) SEM image of the fabricated WDM and (c) representative guided fundamental TE mode profile.}
\end{figure*}

\section{Theory}

Figure~\ref{fig:device} illustrates the inverse-designed WDM geometry and modal operating context. We consider finalized inverse-designed two-channel WDMs with fixed input and output waveguides and a free-form design region represented on a fabrication-aligned grid as a binary layout
\begin{equation}
\bm{\rho} \in \{0,1\}^{H\times W},
\end{equation}
with a minimum feature size of 100~nm.\cite{piggott2015inverse} For various devices produced by the inverse-design pipeline, full-wave simulations provide spectral responses around 1310~nm and 1550~nm. FoMs are defined as transmitted power at these two target wavelengths.

In contrast to the inverse-design optimization stage, we treat the finalized mask $\bm{\rho}$ as a physical object and seek to determine which localized geometric features most strongly govern device performance. The objective is to quantify the sensitivity of transmission metrics to spatially localized perturbations of the layout and to validate these sensitivity predictions through controlled simulation and experimental measurements.

\subsection{Data Generation of Various WDMs}
\begin{figure}[t]
\centering
\includegraphics[width=\columnwidth]{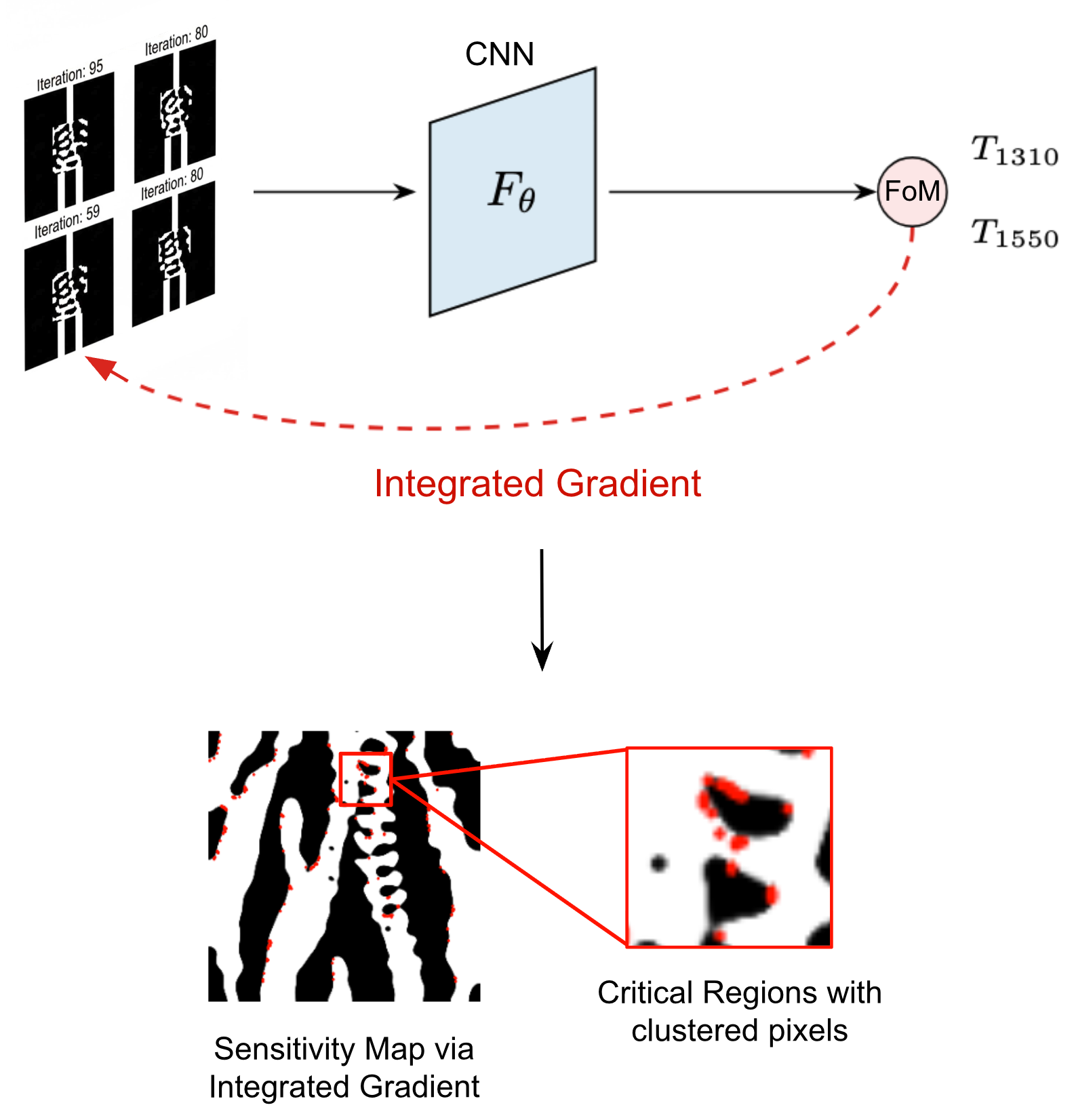}
\caption{\label{fig:pipeline}
Overview of the proposed interpretability workflow. A differentiable surrogate enables pixel-level attribution via Integrated Gradients, producing mask-aligned sensitivity maps and clustered critical regions for analysis and targeted edits.}
\end{figure}

We synthesize a corpus of inverse-designed WDM layouts using SPINS-B with fixed port geometry and randomized seeds and hyperparameters to induce layout diversity with different geometry and FoMs.\cite{SPINSB,su2019nanophotonicinversedesignspins} Each inverse-design run outputs: (i) a finalized binary mask $\bm{\rho}$ and (ii) scalar FoMs. We use a dataset size of \textit{$N = 2200$}. The key requirement is diversity in both geometry and spectral behavior so that the surrogate learns robust layout-to-FoM correlations rather than memorizing a narrow design family. The full interpretability pipeline with the surrogate is summarized in Fig.~\ref{fig:pipeline}.

\begin{figure*}[t]
\centering
\includegraphics[width=0.9\textwidth]{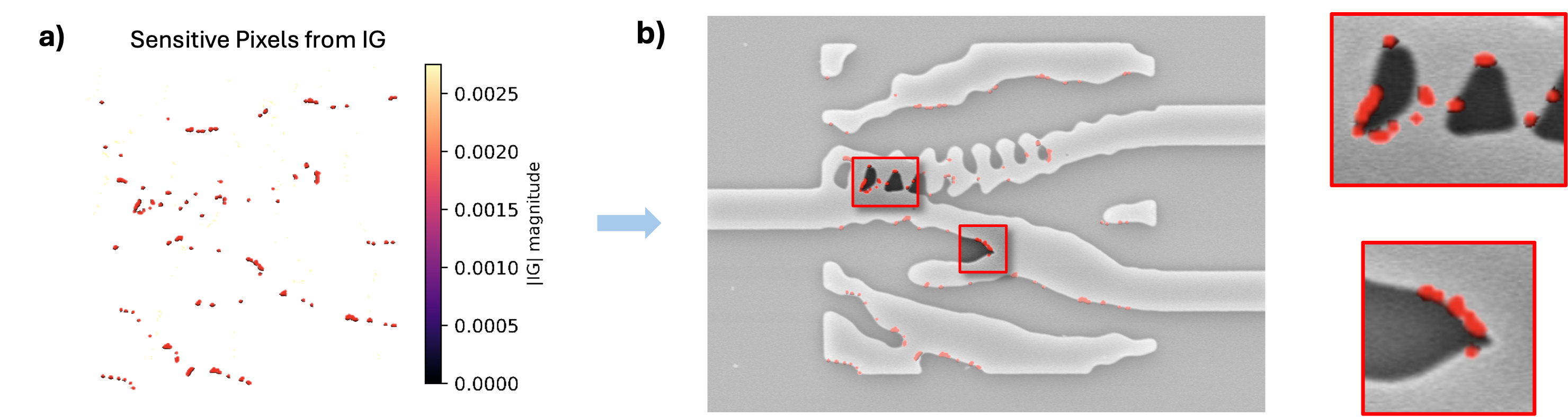}
\caption{\label{fig:igregions}
IG sensitivity overlay and region selection.
(a) Pixel-level IG attribution map (red = high magnitude). (b) Sensitivity map overlaid on SEM image of the unperturbed WDM. Colored boxes denote high-sensitivity regions with clustered pixels.}
\end{figure*}

\subsection{Convolutional Surrogate Model}
We train a lightweight convolutional surrogate $F_{\theta}(\bm{\rho})$ to regress power at 1310~nm and 1550~nm from the binary layout. The training objective is mean-squared error (MSE):
\begin{equation}
\mathcal{L}(\theta)=\frac{1}{N}\sum_{n=1}^{N}\left(F_{\theta}(\bm{\rho}^{(n)})-y^{(n)}\right)^2,
\end{equation}
where $y^{(n)}$ denotes the target FoM for the $n$-th mask.

We optimize using Adam (learning rate $10^{-3}$), with early stopping (patience $=7$) and model selection based on validation MSE. Indices are shuffled with a fixed random seed (42) and split into train/validation/test with a 70/15/15 ratio. The model converges within $\sim$10 epochs and typical MSE values are on the order of $10^{-4}$. We do not claim the surrogate replaces full-wave simulation; rather, it provides a differentiable mapping from mask to metric that enables gradient-based attribution. We validate the interpretability outputs by full-wave re-simulation under controlled perturbations and experimental agreement trends.

\subsection{Pixel-Level Sensitivity Mapping via Integrated Gradients}
To quantify how localized geometric features influence device performance, we compute a spatial sensitivity map directly on the binary layout mask $\bm{\rho}$. We use IG applied to the differentiable surrogate $F_{\theta}(\bm{\rho})$ that predicts transmission metrics. For a given device, IG assigns a contribution value to each pixel,
\begin{equation}
\mathrm{IG}_i = (\rho_i - \rho_{0,i})
\int_{0}^{1}
\frac{\partial F_{\theta}\!\left(\bm{\rho}_0+\alpha(\bm{\rho}-\bm{\rho}_0)\right)}
{\partial \rho_i}
\, d\alpha ,
\end{equation}
which can be interpreted as the accumulated effect of that pixel on the predicted transmission as the layout transitions from a baseline geometry $\bm{\rho}_0$ to the final design.

Because the mask encodes material distribution, $\mathrm{IG}_i$ serves as a proxy for how small, localized geometric perturbations affect modal coupling and interference, thereby providing a pixel-resolved map of performance sensitivity. IG can identify ``high-sensitivity regions'' by grouping clusters of pixels with the largest $|\mathrm{IG}|$ values.

\subsection{Perturbation Methodology: Matched-Area Region Edits}
\begin{figure*}[t]
\centering
\includegraphics[width=\textwidth]{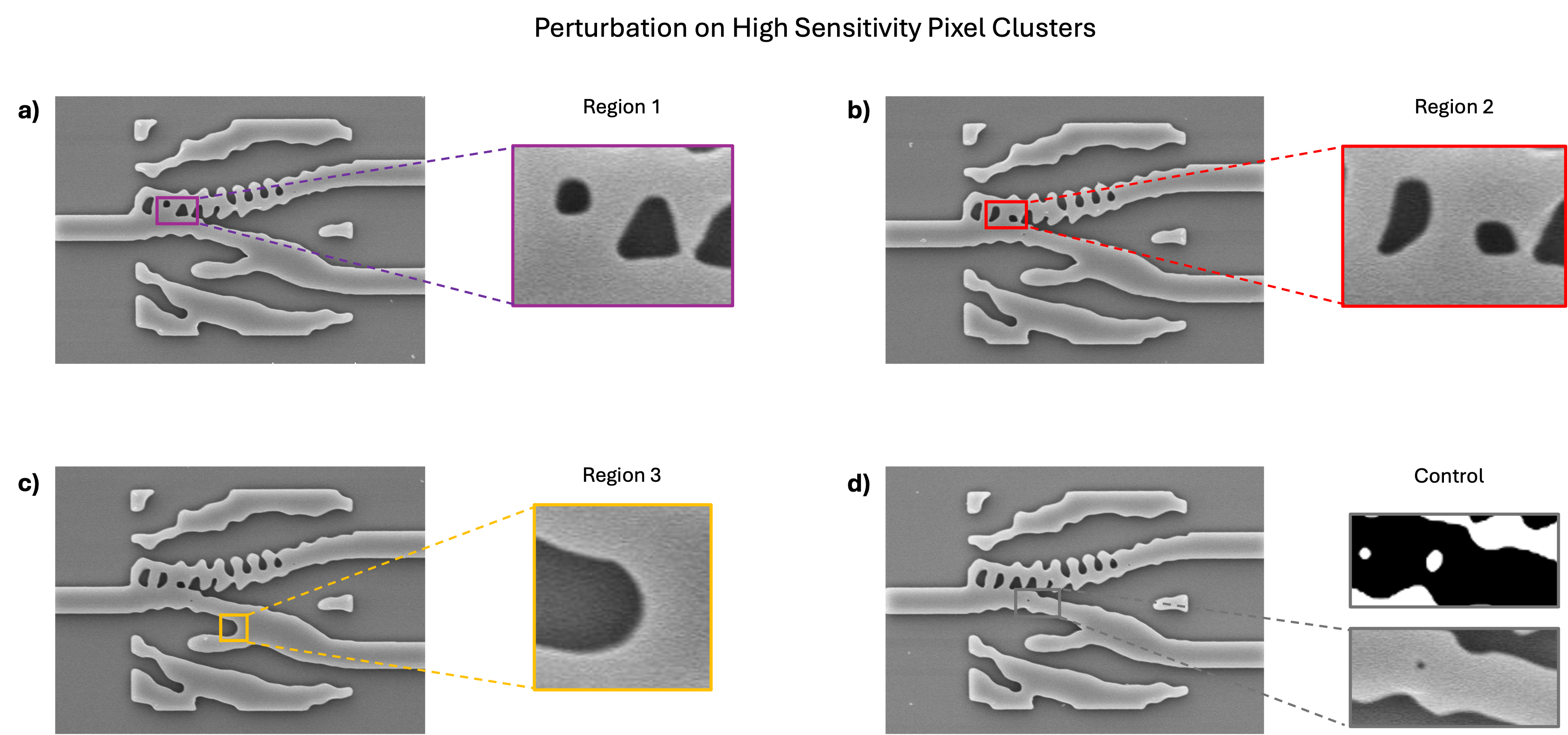}
\caption{\label{fig:perturbs}
SEM visualization of matched-area perturbations applied to IG-identified high-sensitivity regions and a low-sensitivity control.
(a--c) SEM images of fabricated inverse-designed WDMs showing localized fill-in perturbations applied to three high-sensitivity regions (Regions~1--3) identified from clustered Integrated Gradients (IG) attribution hotspots. Insets show magnified views of the perturbed sub-regions.
(d) SEM image of a low-sensitivity control region selected from IG-deprioritized areas, with corresponding mask-level perturbation and fabricated realization shown in the inset.
All perturbations use the same geometric primitive and matched perturbed area ($\sim 0.03~\mu\mathrm{m}^2$), ensuring that observed performance differences arise from \emph{edit location} rather than perturbation magnitude or type.}
\end{figure*}

To validate the IG-derived sensitivity ranking with an apple-to-apple comparison, we generate perturbed masks using the same perturbation primitive and a matched edit budget across selected high-sensitivity regions. High-sensitivity Regions~1--3 are defined by clustered hotspot pixels (top-percentile of $|\mathrm{IG}|$), while a non-sensitive control region is selected from near-zero attribution areas in the IG map (Fig.~\ref{fig:igregions}). We apply a ``fill-in'' perturbation that locally fills etched voids/small notches in the binary mask to emulate lithography/etch bias and corner rounding. For each region, the perturbation is constrained to the same perturbed area ($\sim 0.03~\mu\mathrm{m}^2$; equivalently the same number of 100~nm-grid pixels), ensuring that observed degradation differences arise from \emph{where} the edit is applied rather than perturbation magnitude or type. Fig.~\ref{fig:perturbs} show the perturbed area and the full SEM pictures of the perturbed masks. The same perturbed layouts are used for full-wave simulations and fabrication.

\section{Experiment}

\subsection{Device Fabrication}

Devices were fabricated on a standard silicon-on-insulator (SOI) platform with a 220\,nm-thick silicon device layer on a 2\,$\mu$m-thick buried oxide (BOX) layer. The inverse-designed free-form region contains subwavelength features approaching the 100\,nm mask grid, requiring high-resolution lithography. Electron-beam lithography (EBL) was therefore used with proximity effect correction (PEC) to accurately define fine geometric details. A 300\,nm ZEP520A resist layer was patterned in our in-house EBL with a high-resolution writing mode.

After resist development, an O$_2$ plasma descum was performed to remove residual resist, and the pattern was transferred into the full 220\,nm silicon device layer using reactive ion etching (RIE). Remaining resist was stripped using Nanostrip at 60\,$^\circ$C, and a 1\,$\mu$m Plasma-enhanced Chemical Vapor Deposition SiO$_2$ overcladding was deposited at 350\,$^\circ$C to complete the device stack.

Bias and exposure dose were calibrated to balance minimum achievable feature size with dimensional fidelity. Process optimization targeted faithful reproduction of the splitter/taper hub, high-curvature boundaries, and narrow features identified by the sensitivity analysis.

SEM inspection (Fig.~\ref{fig:device}b) confirms that the free-form region and port interfaces are faithfully transferred, including sub-100~nm features that often dominate sensitivity in inverse-designed layouts.

\subsection{Optical Measurement}
Optical characterization was performed using tunable laser sources covering the O-band and C-band, coupled to the chip via grating couplers. Polarization was adjusted to excite the fundamental TE mode. For each device, transmission into the intended output port was recorded over wavelength sweeps spanning the target bands.

Measured spectra were normalized to a reference waveguide on the same chip to remove coupling and propagation losses external to the device region; nevertheless, residual wavelength-dependent coupling mismatch and chip-scale inhomogeneity can introduce a small systematic offset relative to idealized simulations.

To isolate degradation induced by perturbations, we report excess insertion loss (EIL), defined relative to the unperturbed device in the same band:
\begin{equation}
\mathrm{EIL}_{\mathrm{device}} = \mathrm{IL}_{\mathrm{device}} - \mathrm{IL}_{\mathrm{Unperturbed}}.
\label{eq:eil}
\end{equation}
This normalization suppresses common-mode offsets between experiment and simulation and directly quantifies the penalty attributable to localized edits.

\section{Results}

\begin{figure*}[!t]
\centering
\includegraphics[width=\textwidth]{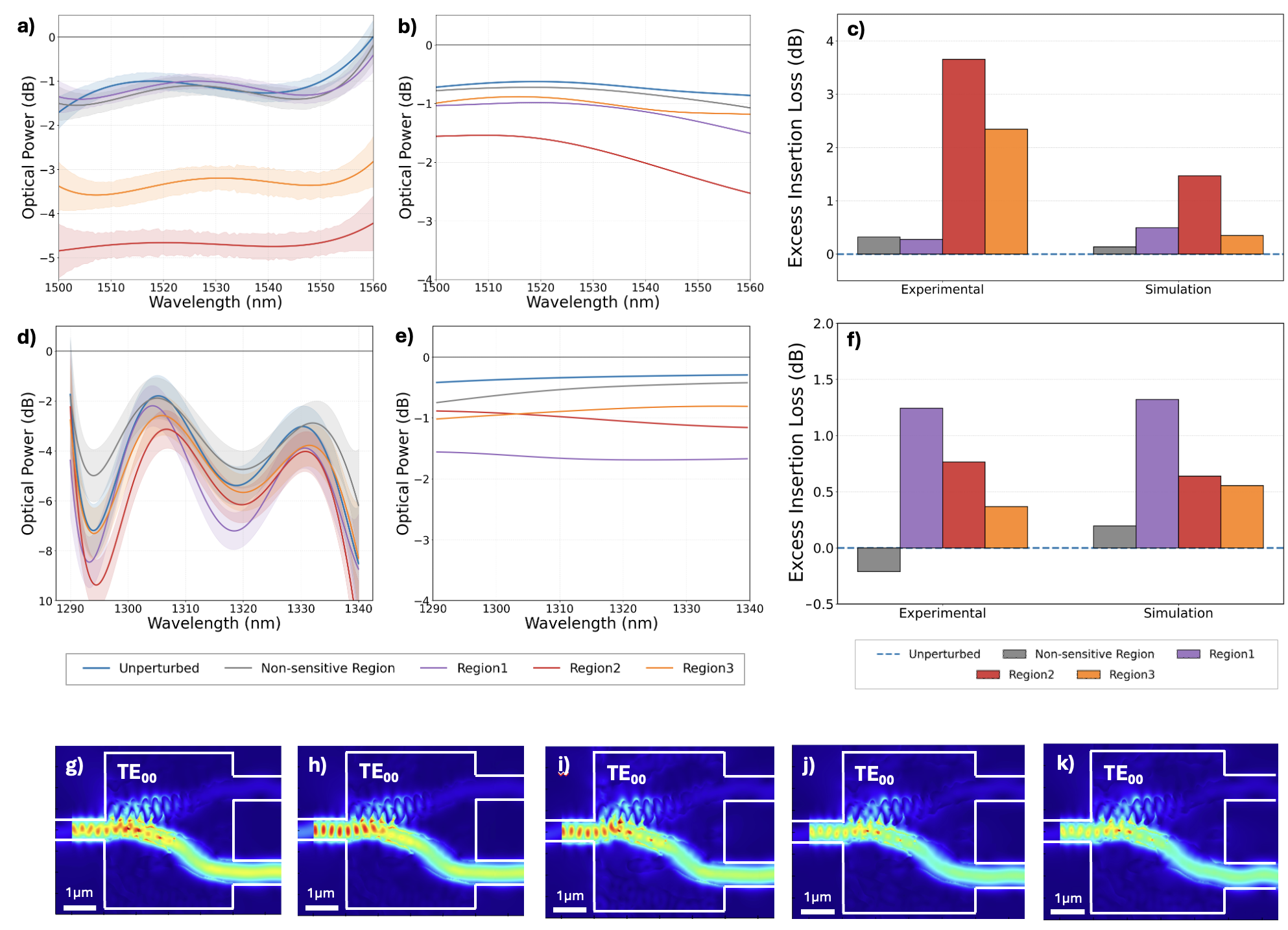}
\caption{\label{fig:simperturb}
Experimental and simulated validation of sensitivity-guided perturbations, with corresponding mode-propagation analysis.
(a) Measured transmission spectra near 1550~nm (C-band) for the unperturbed device, low-sensitivity control, and high-sensitivity region edits.
(b) Corresponding full-wave simulated transmission at 1550~nm.
(c) Excess insertion loss (EIL) at 1550~nm comparing experiment and simulation, referenced to the unperturbed device.
(d) Measured transmission spectra near 1310~nm (O-band).
(e) Corresponding full-wave simulated transmission at 1310~nm.
(f) EIL at 1310~nm for experiment and simulation.
(g--k) Simulated electric-field intensity $|E|^2$ distributions of the fundamental TE$_{00}$ mode at the O-band (1310~nm) for, in order, the unperturbed device, low-sensitivity control, and high-sensitivity perturbations applied to Regions~1, 2, and~3.}
\end{figure*}

We first use IG to identify critical geometric substructures and define matched-area perturbation regions (Fig.~\ref{fig:igregions}). We then implement the corresponding localized ``fill-in'' edits in fabrication; Fig.~\ref{fig:perturbs} provides SEM verification of the matched-area perturbations for Regions~1--3 and the low-sensitivity control. Finally, we compare measured and simulated spectra and excess insertion loss to validate that IG-identified hotspots correspond to true fabrication-sensitive regions (Fig.~\ref{fig:simperturb} and Table~\ref{tab:summary}).


\subsection{Sensitivity Maps}
Pixel-level IG attribution produces spatially sparse sensitivity maps in which a limited subset of geometric features dominate device performance (Fig.~\ref{fig:igregions}). Hotspots consistently localize to high-curvature dielectric boundaries, abrupt width transitions, and the central splitter/taper hub responsible for modal conversion and phase partitioning. These regions align with known electromagnetic sensitivity mechanisms in silicon photonics~\cite{wang2011robust}: curvature and corners enhance scattering loss, abrupt transitions introduce impedance discontinuities, and the hub controls modal superposition and relative phase that downstream geometry converts into wavelength-selective routing~\cite{payne1994theoretical}.

This localization indicates that performance is governed by a small number of ``structural control points,'' while large portions of the free-form region exhibit near-zero attribution. Once the interference scaffold is established, many pixels become functionally redundant with respect to the evaluated FoMs, consistent with interference-based device physics.

\begin{table*}[t]
\centering
\caption{\label{tab:summary}
Summary of experimental and simulated performance. For both 1550~nm and 1310~nm bands, insertion loss (IL) and excess insertion loss (EIL) are referenced to the unperturbed device in the same band. Exp denotes experimental and Sim denotes simulation.}
\setlength{\tabcolsep}{6.5pt}
\renewcommand{\arraystretch}{1.25}
\small
\begin{ruledtabular}
\begin{tabular}{lcccccccc}
Device & \multicolumn{4}{c}{1550~nm (C-band)} & \multicolumn{4}{c}{1310~nm (O-band)} \\
& Exp IL & Exp EIL & Sim IL & Sim EIL & Exp IL & Exp EIL & Sim IL & Sim EIL \\
\hline
Unperturbed & 0.595 & 0.000 & 0.812 & 0.000 & 2.9058 & 0.0000  & 0.3395 & 0.0000 \\
Non-sensitive & 0.915 & 0.320 & 0.946 & 0.134 & 2.6935 & -0.2123 & 0.5354 & 0.1960 \\
Region1 & 0.869 & 0.274 & 1.306 & 0.494 & 4.1491 & 1.2433  & 1.6609 & 1.3215 \\
Region2 & 4.248 & 3.653 & 2.282 & 1.470 & 3.6694 & 0.7636  & 0.9781 & 0.6386 \\
Region3 & 2.939 & 2.344 & 1.161 & 0.349 & 3.2737 & 0.3679  & 0.8939 & 0.5544 \\
\end{tabular}
\end{ruledtabular}
\end{table*}

\subsection{FDTD Simulation}
As mentioned above, matched-area fill-in perturbations were applied to both control and IG-identified high-sensitivity regions (SEM examples shown in Fig.~\ref{fig:perturbs}), followed by full-wave FDTD re-simulation. Because perturbation type and area are fixed, degradation differences isolate the effect of \emph{where} the geometry is modified.

Simulations show a clear quantitative separation (Fig.~\ref{fig:simperturb}c and Fig.~\ref{fig:simperturb}f, and Table~\ref{tab:summary}): high-sensitivity edits induce substantially larger EIL than control edits under identical perturbation budgets and produce positive EIL from the unperturbed baseline. At 1550~nm, simulated EIL increases from 0.134~dB (control) to 1.470~dB (Region~2), an $\sim11\times$ increase. At 1310~nm, Region~1 produces 1.3215~dB EIL compared with 0.1960~dB for the control.

Mode-propagation analysis further clarifies the physical origin of this degradation (Fig.~\ref{fig:simperturb}g--k). Perturbations in Regions~1 and~2 primarily reduce the transmitted $|E|^2$ magnitude at the output port while preserving the fundamental TE$_{00}$ modal profile, consistent with increased scattering and phase imbalance. In contrast, the Region~3 perturbation (Fig.~\ref{fig:simperturb}k) induces pronounced spatial oscillations near the output port, indicating unstable modal interference and partial excitation of non-ideal field components. These observations confirm that attribution hotspots correspond to locations where small geometric changes strongly perturb modal interference and phase balance, ultimately degrading the device FoM.

\subsection{Experimental Validation}
Measured performance (Fig.~\ref{fig:simperturb} and Table~\ref{tab:summary}) preserves the simulation-predicted hierarchy. At 1550~nm, EIL rises from 0.320~dB (control) to 2.344~dB (Region~3) and 3.653~dB (Region~2), corresponding to an $\sim11.4\times$ larger penalty for Region~2. At 1310~nm, Region~1 exhibits the largest EIL (1.2433~dB), followed by Region~2 (0.7636~dB) and Region~3 (0.3679~dB), while the control shows a small negative EIL consistent with normalization uncertainty.

\textbf{Simulation--experiment comparison.}
The experimental results preserve the simulation-predicted sensitivity hierarchy: regions identified as high-sensitivity in simulation consistently exhibit the largest excess insertion loss in fabricated devices, while control regions show minimal degradation. Experimental penalties, however, exceed simulated values at the most sensitive 1550~nm hotspots. This amplification is consistent with fabrication-induced scattering (sidewall and line-edge roughness, etch microloading) not captured in nominal-geometry FDTD, as well as nonlinear EBL/etch bias distortions that effectively increase perturbation severity in interference-critical regions. Grating-coupler reflections introduce additional Fabry--P\'erot-like effects not included in simulation, further enhancing apparent loss~\cite{wang2014focusing}. At 1310~nm, Region~1 shows close agreement (1.2433 vs.~1.3215~dB), indicating degradation dominated by mode mismatch and phase error captured by the solver.

\section{Discussion}
The spatial localization of hotspots aligns with established electromagnetic sensitivity mechanisms. High-curvature boundaries act as scattering centers, abrupt transitions introduce impedance discontinuities, and the splitter/taper hub controls modal phase partition. Perturbations in these regions disrupt interference balance, explaining the disproportionately large excess loss observed experimentally.

Free-form inverse-designed layouts lack the geometry-localized tolerance abstractions typical of PDK cells. The demonstrated sensitivity mapping translates complex masks into actionable manufacturing priorities. Hotspot-aware constraint allocation can enforce tighter feature rules only where performance is most sensitive, while sensitivity overlays guide metrology across the entire manufacturing flow, from mask reticle fabrication to wafer processing, to post-fabrication inspection, and ultimately failure analysis. Ranking geometry by functional importance supports reduced-order abstraction of free-form layouts, facilitating integration into PDK-style flows.

Sensitivity estimates derive from gradients of a learned surrogate rather than direct field-overlap sensitivities. Agreement between attribution trends, full-wave perturbation simulations, and measurements indicates that dominant geometry--performance relationships are captured, though fabrication-induced scattering and measurement parasitics can amplify degradation beyond nominal simulations. Future work may incorporate fabrication-aware perturbations and uncertainty-aware surrogates to improve quantitative prediction.

\section{Conclusion}
We demonstrated an experimentally validated framework for deriving pixel-level fabrication sensitivity maps from inverse-designed photonic layouts. The resulting attributions localize to electromagnetically critical substructures, and matched-area perturbation tests show that edits in these regions produce substantially larger excess insertion loss than geometrically equivalent edits in low-sensitivity areas. Agreement between simulation and fabricated-device measurements establishes that the learned sensitivity map corresponds to true interference- and fabrication-relevant device physics. Beyond interpretability, this approach provides a practical pathway toward tolerance-aware constraint allocation, targeted metrology, and manufacturability-informed abstraction of free-form inverse-designed components, supporting their integration into foundry-compatible photonic design workflows.

\begin{acknowledgments}
The experimental portion of this work was performed in part at the University of Michigan Lurie Nanofabrication Facility. We also thank the Stanford NQP group for releasing SPINS-B used in this work.
\end{acknowledgments}

\section*{Data Availability}
The data that support the findings of this study are available from the corresponding author upon request.

\bibliographystyle{aipnum4-1}
\bibliography{aipsamp}

\end{document}